\documentclass[letterpaper, 10pt, conference]{ieeeconf}

\usepackage{times}
\usepackage{tabularx}
\usepackage{multirow,multicol, array}
\usepackage{graphicx}
\usepackage{wrapfig}
\usepackage{nccmath}
\usepackage{hyperref}

\usepackage{amsmath,amssymb,amsfonts}
\usepackage[ruled,vlined,linesnumbered]{algorithm2e}
\usepackage{textcomp}
\usepackage{acronym}
\usepackage{balance}
\usepackage{mdwmath}
\usepackage{bm}
\usepackage{mdwtab}
\usepackage{array}
\usepackage{float}

\usepackage[usenames,dvipsnames]{color}
\usepackage{eqparbox}
\usepackage{cite}

\usepackage{color}
\usepackage{psfrag}
\usepackage{epsfig}
\usepackage{url}

\usepackage{epstopdf}
\usepackage{booktabs}
\usepackage{blindtext}

\usepackage[colorinlistoftodos,prependcaption,textsize=tiny]{todonotes}

\IEEEoverridecommandlockouts
\begin{document}


\title{A Data-driven Hierarchical Control Structure for Systems with Uncertainty}

\author{Lu Shi, Hanzhe Teng, Xinyue Kan, and 
Konstantinos Karydis
\thanks{The authors are with the Dept. of Electrical and Computer Engineering, University of California, Riverside. 
	Email: {\{lshi024, hteng007, xkan001, 
	karydis\}@ucr.edu}.}
\thanks{We gratefully acknowledge the support of NSF under grants \#IIS-1910087 and \#IIS-1724341, and ONR under grant \#N00014-19-1-2264. Any opinions, findings, and conclusions or recommendations expressed in this material are those of the authors and do not necessarily reflect the views of the National Science Foundation.}
}

\maketitle
\thispagestyle{empty}


\begin{abstract}
    The paper introduces a Data-driven Hierarchical Control (DHC) structure to improve performance of systems operating under the effect of system and/or environment uncertainty. The proposed hierarchical approach consists of two parts: 1) A data-driven model identification component to learn a linear approximation between reference signals given to an existing lower-level controller and uncertain time-varying plant outputs. 2) A higher-level controller component that utilizes the identified approximation and wraps around the existing controller for the system to handle modeling errors and environment uncertainties during system deployment.  
    We derive loose and tight bounds for the identified approximation's sensitivity to noisy data. Further, we show that adding the higher-level controller maintains the original system's stability. A benefit of the proposed approach is that it requires only a small amount of observations on states and inputs, and it thus works online; that feature makes our approach appealing to robotics applications where real-time operation is critical. The efficacy of the DHC structure is demonstrated in simulation and is validated experimentally using aerial robots with approximately-known mass and moment of inertia parameters and that operate under the influence of ground effect. 
\end{abstract}


\section{Introduction}



As robots increasingly venture outside of the lab, the effect of uncertainty within the system and/or at the robot-environment interactions becomes more pronounced. Motivating examples include the influence of surface effects for Unmanned Aerial Vehicles (UAVs), unknown flow dynamics for marine robots, and inherently uncertain leg-ground contacts for legged robots. 

Central to exploiting system and robot-environment uncertainty is the ability to quickly identify deviations from nominal behaviors based on data collected during robot deployment. Pre-deployment model identification and calibration tools (e.g.,~\cite{c30,c18,c20}) can help improve 
model-based control. However, even though such models can be obtained with high precision, the presence of uncertain, time-varying disturbances during deployment may turn the utilized model invalid to the extent that an otherwise well-tuned controller will fail~\cite{c37}.  
A way to tackle this is to incorporate Uncertainty Quantification (UQ) into control design.

Methods using UQ, like a-priori estimation, rely heavily on the employed model structure. Generated models are usually complex and hard to incorporate into controller design in practice~\cite{c12,c24}. 
When prior information about the environment is limited, methods based on principles of 
adaptive control~\cite{c22} can apply. An appealing feature of these methods is that they can provide some form of performance and safety guarantees (e.g., regarding system stability). 
Despite their overall effectiveness, 
such methods can still be challenged in two key ways. 
First, when the intensity of the uncertainty affecting system behavior violates the underlying controller assumptions. 
Second, when modeling errors~\cite{c23}, such as unmodeled dynamics, render the model invalid~\cite{c37}. To address the problem of unmodeled dynamics, data-driven control techniques have been proposed. 

Machine learning methods, such as Gaussian Process~\cite{c9,c10}, or Deep Neural Networks~\cite{c16}, can be used to either model the system~\cite{c39,c38} or determine control inputs \cite{c43,c44} following a black-box input-output training procedure. 
However, as state dimensionality and system structure complexity increase (e.g., by using `deeper' neural networks), the aforementioned methods may be challenged when it comes to be implemented in real-time for robotics research (e.g.,~\cite{PP2017offline,emran2017offline,claire2016offline}).
One way to address this issue is by adopting a hierarchical structure to involve deep neural networks in control design.  Instead of designing new controllers directly, reference trajectories are generated to deal with uncertainty through learning~\cite{c41,APhierarchical}. 
Unfortunately, certain limitations still exist even when adopting a hierarchical structure. Neural-network-based approaches require a large body of data to train well~\cite{AP2019histability}. At the same time, they remain limited in terms of offering some form of performance and safety guarantees, although this is currently an active research topic (e.g.,~\cite{CJ2019hier,APhierarchical, EL2018hier}).

Besides neural-network-based approaches, dimensionality reduction and dynamics decomposition approaches can play an important role in data-driven algorithms. Methods like DMD (Dynamic Mode Decomposition~\cite{DMD}), EDMD (Extended DMD~\cite{EDMD}), POD (Proper Orthogonal Decomposition~\cite{POD,kPOD}) and their various kernel~\cite{kernel} and tensor~\cite{c8} extensions have been successfully applied to across areas. A benefit of decomposition algorithms is that they can significantly reduce the amount of data required for approximating a system's model through data (e.g.,~\cite{TD2019dmd,JN2016dmd,cortez2019dmd}). 

\begin{figure}
\vspace{6pt}
\centering
\includegraphics[trim= 0.1cm 0cm 0.1cm 0.3cm, clip,width =0.47\textwidth]{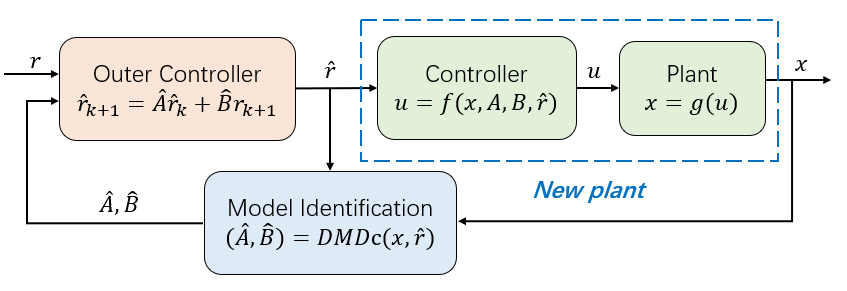}
\vspace{-9pt}
\caption{Overview of the Data-driven Hierarchical Control (DHC) structure proposed in this work. Model identification (via spectral methods) is combined with a linear controller to adjust the reference input given to an existing (pre-tuned) lower-level controller.}
\label{fig:MiC_cascade}
\vspace{-12pt}
\end{figure}

Fueled by the potential of hierarchical methods and dimensionality-reduction approaches, this paper presents a new \underline{D}ata-driven \underline{H}ierarchical \underline{C}ontrol (DHC) structure to handle uncertainty.  The approach hinges on DMD with Control (DMDc~\cite{DMDc}) to learn a higher-level controller and then generate a refined reference to improve an existing lower-level controller's performance (Fig.~\ref{fig:MiC_cascade}).
Our approach can be particularly appropriate in practice when a low-level, high-rate, pre-tuned controller is already in place, and a second higher-level controller is wrapped around the lower-level one to allow for the system to operate under uncertainty. The quadrotor UAV is one illustrative case, where it is now common practice to employ a hierarchical control structure~\cite{kumar2012mav, mellinger2010mav}. 
The low-level controller is responsible for controlling the attitude of the robot, while the high-level controller determines its position. 

Succinctly, the paper's contribution is twofold.
\begin{itemize}
    \item We propose a hierarchical control structure that refines a reference signal sent to a lower-level controller to deal with uncertainty. The structure builds on top of a model-identification block based on DMDc, and is shown that it retains stability of the underlying controller.
    \item We analyze the sensitivity of DMDc to noisy data, and provide loose and tight bounds for the model identification component of our method.
\end{itemize}
Furthermore, we evaluate and validate the methodology using aerial robots both in simulation and experimentally. In turn, this effort can help design controllers able to learn how to harness uncertain aerodynamics, such as ground effect. A supplementary video can be found at \url{https://youtu.be/OznDCskVnJU}.


\section{Technical Background}\label{sec:analysis}
Dynamic Mode Decomposition (DMD) can characterize nonlinear dynamics through analysis of an approximated linear system~\cite{DMDgeneral}. 
Most relevant to our work here, DMD with control (DMDc)--an extension of the original DMD~\cite{DMDfirst}---aims to incorporate the effect of external inputs~\cite{DMDc}. 

The goal of DMDc is to analyze the relation between a future system measurement $\xi_{k+1} \in \mathbb{R}^n$, a current measurement $\xi_k \in \mathbb{R}^n$, and the current input $u_k \in \mathbb{R}^m$. 
For each triplet of measurement data $(\xi_{k+1}, \xi_{k}, u_{k})$, a pair of linear operators 
$A \in \mathbb{R}^{n\times n}$, and $B \in \mathbb{R}^{n\times m}$ is determined to approximate 
\begin{equation}\label{eq:approx}
\xi_{k+1} \approx A\xi_k+Bu_k \enspace.
\end{equation}

Operators $A$ and $B$ are selected as the best-fit solution for all triplets of available data. 
Given observations and inputs up to time instant $M$ arranged in vectors $\Xi = [\xi_1,\xi_2 \dots, \xi_{M-1}]$, $\Xi_P = [\xi_2,\xi_3\dots, \xi_M]$, and $U = [u_1,u_2\dots, u_{M-1}]$, 
approximation~\eqref{eq:approx} can be rewritten compactly as
\begin{equation}\label{eq:DMDc}
    \left[
          \begin{matrix}
            A&B\\
          \end{matrix} 
          \right]\left[
          \begin{matrix}
            \Xi\\
           U \\
          \end{matrix} 
    \right] = A_P\Omega^T\approx \Xi_P \enspace.
\end{equation}
Then we could seek the best-fit solution as:
$$A_P = \Xi_P(\Omega^T)^{\dagger}\enspace ,$$
where $\dagger $ denotes the pseudo-inverse. The problem can be solved immediately by Singular Value Decomposition (SVD), or QR decomposition, among others~\cite{c1,c2}.

\section{Data-driven Hierarchical Control (DHC)}\label{sec:structure}
\subsection{Controller Structure}\label{ssec:overview}
Given an existing low-level controller for a plant of interest, the proposed DHC structure adds a model identification block and a higher-level (``outer") controller to extract uncertainties and remedy for them in real-time, respectively (Fig.~\ref{fig:MiC_cascade}). The low-level controller and the original plant are combined into a new plant. The model identification block is built from DMDc and estimates linear operators $\hat{A}$ and $\hat{B}$ from new plant output and reference signals. The identified operators are then used by the higher-level controller to refine the reference signal sent to the new plant.

\subsection{Control Procedure}\label{ssec:process}
At timestep $k+1$, the outer controller refines the original reference $r_{k+1}$ given to the lower-level controller as per $\hat{r}_{k+1} = \hat{A}_{k}\hat{r}_{k}+\hat{B}_{k}r_{k+1}$. 
To estimate $\hat{A}_k$ and $\hat{B}_k$ using DMDc, we treat the measured state $x_k$ as the ``input" and the refined references $\hat{r}_k$ and $\hat{r}_{k+1}$ as the evolving ``output." The reason for doing so is that the controller operates on the reference signal, which in turn needs to be ``learned" based on observed system state. 
Then, DMDc operates on the linear approximation 
$\hat{r}_{k+1} = \hat{A}_{k}\hat{r}_{k}+\hat{B}_{k}x_{k}$. 
%
%
Note that to avoid singularities, we check the rank of the measurement matrices used to approximate models. 
After an initialization for $M$ steps,\footnote{From a general standpoint,~\eqref{eq:DMDc} can be solved when $m+n+1$ linearly independent measurements are available~\cite{DMDc}. Thus, a lower bound for $M$ is $m+n+1$. In this work we set $M=m+n+1$.} the identification process repeats to refine the estimated model until the control task is finished (Algorithm~\ref{alg}). 



\vspace{-6pt}
\begin{algorithm}[h]
\caption{DHC Procedure}\label{alg}
\textbf{initialize:} Set $\hat{r}_i = r_i$ and evolve the system for the first $M$ time steps. 
Then compute
$(\hat{A}_{0},\hat{B}_{0}) = DMDc(\left\lbrace \hat{r}_{i}\right\rbrace_{i=1}^{M},\left\lbrace x_{i}\right\rbrace_{i=1}^{M})$
\\
\For{$k\geq M$}{
\Repeat{\text{control task is finished}}{
\textit{\textbf{Outer Controller}}:  Refine reference based on approximated model: 
            $\hat{r}_{k+1} = \hat{A}_{k}\hat{r}_{k}+\hat{B}_{k}r_{k+1}$.
        \\
\textit{\textbf{Plant}}: Propagate the system with $\hat{r}_{k+1}$:
    $
        x_{k+1} = g(\hat{r}_{k+1},w_{k})
    $
    , where $g$ is the evolving law of real plant and $w_{k}$ the uncertainty.
\\
\textit{\textbf{Model}}: Check rank of measurement matrices. ~ \If{full column rank}{
$(\hat{A}_{k},\hat{B}_{k}) = DMDc(\left\lbrace \hat{r}_{i}\right\rbrace^{k+1},\left\lbrace x_{i}\right\rbrace^{k+1})$
} ~~~~~
\Else
{$(\hat{A}_{k},\hat{B}_{k}) = (\hat{A}_{k-1},\hat{B}_{k-1})$}
\textbf{$k \gets k+1$}
}
}
\end{algorithm}
\vspace{-9pt}

\subsection{Stability Analysis}
Let the underlying low-level controller be stable, and 
consider the origin as the reference, i.e. $r_k = 0$. Given the Lyapunov function
$V(x_k) = x_k^Tx_k$, we calculate $\Delta V(x_k) = V(x_{k+1})-V(x_k)=g(\hat{r}_{k+1}, w_{k+1})^Tg(\hat{r}_{k+1}, w_{k+1})-x_k^Tx_k$. 
When $\hat{r}_{k+1} = \hat{A}_k\hat{r}_k+\hat{B}_kr_{k+1}$, we have  $g(\hat{r}_{k+1}, w_{k+1}) = r_{k+1} = 0$. Then, $\Delta V(x_k) =  r_{k+1}^T r_{k+1}-x_k^Tx_k = 0-x_k^Tx_k
\to \Delta V(x_k) = -x_k^Tx_k\enspace.$

Thus, $V$ decays over time and satisfies $V(0) = 0$. From the standard Lyapunov argument, the equilibrium $r_k = 0$ is then stable. Hence, wrapping the outer controller around the underlying plant respects the stability properties of the original lower-level controller.






\section{Sensitivity to Noisy Data}\label{sec:sensitivity}
The model identification block extracts dynamics when measurement data are noisy. Below we analyze the sensitivity to noisy data and derive two estimation error bounds. 

    \textbf{Lemma 2.1.} 
    Consider the system  
\begin{equation}\label{eq:error}
   \begin{array}{rll}{H\textbf{z}} & {=\textbf{s}\enspace,}  & {H \in \mathbb{R}^{p \times q},\enspace  \textbf{s} \in \mathbb{R}^{p \times t}} \\ {(H+\delta H) \hat{\textbf{z}}} & {=\textbf{s}+\delta \textbf{s}\enspace,} & {\delta H \in \mathbb{R}^{p \times q},\enspace \delta \textbf{s} \in \mathbb{R}^{p \times t}}\end{array}
\end{equation}
%
%
The sensitivity of system~\eqref{eq:error} to perturbations in $\textbf{s}$ and $H$ is
\begin{equation}\label{eq:loose_bound}
\begin{medsize}
    \frac{\|\hat{\textbf{z}}-\textbf{z}\|}{\|\textbf{z}\|} \leq\left(\kappa(H)^{2} \tan (\theta)+\kappa(H)\right) \frac{\|\delta H\|}{\|H\|}+\kappa(H) \sec (\theta) \frac{\|\delta \textbf{s}\|}{\|\textbf{s}\|}\enspace.
\end{medsize}
\end{equation}
where $\kappa(H)$ is the condition number of linear operator H, and $\theta=\arccos{\frac{\|H \textbf{z}\|}{\|\textbf{s}\|}}.$

\vspace{2pt}
\textit{Proof of Lemma 2.1:}
Start with the normal equations
$$H^TH\textbf{z} = H^T\textbf{s}\enspace.$$
The first-order perturbation relation is
$$\delta H^TH\textbf{z}+H^T\delta H\textbf{z}+H^TH\delta\textbf{z} = \delta H^T\textbf{s}+H^T\delta\textbf{s}\enspace,$$
which we re-arrange to get
$$\delta \textbf{z} = (H^TH)^{-1}\delta H^T(\textbf{s}-H\textbf{z})+(H^TH)^{-1}H^T(\delta \textbf{s}-\delta H \textbf{z}).$$
Define $r = \textbf{s}-H\textbf{z}$, and let $\sigma_1,\dots,\sigma_p$ be singular values of $H$. Then, 
%

\begin{equation*}
    \begin{split}
        \|\delta \textbf{z}\| \leq & \|(H^TH)^{-1}\| \|\delta H\|\|r\| \\
        &+\|(H^TH)^{-1}H^T\|\left(\|\delta \textbf{s}\|+\|\delta H\|\|\textbf{z}\|\right) \enspace.
    \end{split}
\end{equation*}
%
%
Setting $\|H\| = \sigma_1$, $\|(H^TH)^{-1}\| = 1/\sigma_q^2$, we can derive~\cite{c3} 
$$\|\delta \textbf{z}\| \leq \frac{\|\delta H\|}{\sigma_{q}^{2}}\|r\|+\frac{1}{\sigma_{q}}(\|\delta \textbf{s}\|+\|\delta H\|\|\textbf{z}\|) \enspace .$$

Dividing both sides by $\|\textbf{z}\|$ and after some algebraic manipulation so that $\|\delta H\|$ and $\|\delta \textbf{s}\|$ only appear in ratios of $\frac{\|\delta H\|}{\| H\|}$ and $\frac{\|\delta \textbf{s}\|}{\|\textbf{s}\|}$, as $\kappa(H) = \sigma_1/\sigma_q$, we get 
\begin{equation*}
 \begin{medsize}
    \frac{\|\delta \textbf{z} \|}{\|\textbf{z}\|} \leq \kappa(H)^{2} \frac{\|r\|}{\|H\|\|\textbf{z}\|} \frac{\|\delta H\|}{\|H\|}+\kappa(H)\left(\frac{\|\textbf{s}\|}{\|H\|\|\textbf{z}\|} \frac{\|\delta \textbf{s}\|}{\|\textbf{s}\|}+\frac{\|\delta H\|}{\|H\|}\right)\enspace.
\end{medsize}   
\end{equation*}
Then with
$$
\begin{array}{l}{\frac{\|r\|}{\|H\|\|\textbf{z}\|} \leq \frac{\|r\|}{\|H \textbf{z}\|}=\tan (\theta)\enspace,} \\ {\frac{\|\textbf{s}\|}{\|H\|\|\textbf{z}\|} \leq \frac{\|\textbf{s}\|}{\|H \textbf{z}\|}=\sec (\theta)\enspace,}\end{array} 
$$
we arrive at \eqref{eq:loose_bound}.$\hfill\blacksquare$

\textbf{Corollary 2.2.} The estimation error of problem~\eqref{eq:DMDc} is bounded by the measurements' noise intensity, i.e. 
\begin{equation}\label{eq:loose}
\begin{medsize}
    \frac{\|\delta A_P\|}{\|A_P\|} \leq \left(\kappa(\Omega)^{2} \tan (\theta)+\kappa(\Omega)\right) \frac{\|\delta \Omega\|}{\|\Omega\|}+\kappa(\Omega) \sec (\theta) \frac{\|\delta \Xi_P\|}{\|\Xi_P\|},
\end{medsize}
\end{equation}
where $\kappa(\Omega)$ is the condition number of observation matrix $\Omega$, and $\theta=\arccos{\frac{\|A_P\Omega \|}{\|\Xi_P\|}}.$

\vspace{2pt}
\textit{Proof of Corollary 2.2:} The proof follows directly from Lemma 2.1 by setting $H = \Omega$, $\textbf{z} = A_P^T$, and $\textbf{s} = \Xi_P^T$. 
$\hfill\blacksquare$

\vspace{2pt}
\textbf{Lemma 2.3 [A tighter bound].} Let $\epsilon = \frac{\|\delta \textbf{s}\|}{\|\textbf{s}\|}$. In the case that $\|\delta \textbf{s}\|  =  \|\delta H\|$, the sensitivity of \eqref{eq:error} is reduced to 
\begin{equation}\label{eq:tight_bound}
    \frac{\|\hat{\textbf{z}}-\textbf{z}\|}{\|\textbf{z}\|} \leq \epsilon \kappa(H)(1+\|\hat{\textbf{z}}\|)\enspace .
\end{equation}

\vspace{2pt}
\textit{Proof of Lemma 2.3:} From system~\eqref{eq:error}, we have 
$$H(\hat{\textbf{z}}-\textbf{z})=\delta \textbf{s}- \delta H \hat{\textbf{z}}\enspace ,$$
and it follows that 
$$
\hat{\textbf{z}}-\textbf{z}=(H^TH)^{-1}H^T \delta \textbf{s}-(H^TH)^{-1}H^T \delta H \hat{\textbf{z}}\enspace .
$$
Then, 
$$
\|\hat{\textbf{z}}-\textbf{z}\|=\|(H^TH)^{-1}H^T\| (\|\delta \textbf{s}\|+\|\delta H\|\| \hat{\textbf{z}}\|)\enspace .
$$
Setting $\|\delta \textbf{s}\|  =  \|\delta H\|$ yields 
\begin{equation*}
    \begin{split}
        \|\hat{\textbf{z}}-\textbf{z}\| &= \|(H^TH)^{-1}H^T\| \|\delta \textbf{s}\|(1+\| \hat{\textbf{z}}\|) \\ 
        &= \epsilon\|(H^TH)^{-1}H^T\| \| \textbf{s}\|(1+\| \hat{\textbf{z}}\|) \\ 
        &\leq \epsilon \|(H^TH)^{-1}H^T\| \|H\|\| \textbf{z}\|(1+\| \hat{\textbf{z}}\|) \\ 
        &=\epsilon \kappa(H)\|\textbf{z}\|(1+\| \hat{\textbf{z}}\|)\enspace,
    \end{split}
\end{equation*}
which concludes the proof. $\hfill\blacksquare$

\textbf{Corollary 2.4.} When the disturbance occurs in observed states, i.e. $\|\delta \Xi_P\| =\|\delta \Xi\|=\|\delta \Omega\|$ a tighter error bound is
\begin{equation}\label{eq:tight}
    \frac{\|\delta A_P\|}{\|A_P\|} \leq \kappa(\Omega) \frac{\|\delta \Xi_p\|}{\|\Xi_p\|} (1+\|\hat{A}_p\|)\enspace .
\end{equation}

\textit{Proof of Corollary 2.4:} The proof follows directly from Lemma 2.3 by setting $H = \Omega$, $\textbf{z} = A_P^T$, and $\textbf{s} = \Xi_P^T$. 
$\hfill\blacksquare$

\begin{figure}[!ht]
\vspace{0pt}
\centering
\includegraphics[width = 0.35\textwidth]{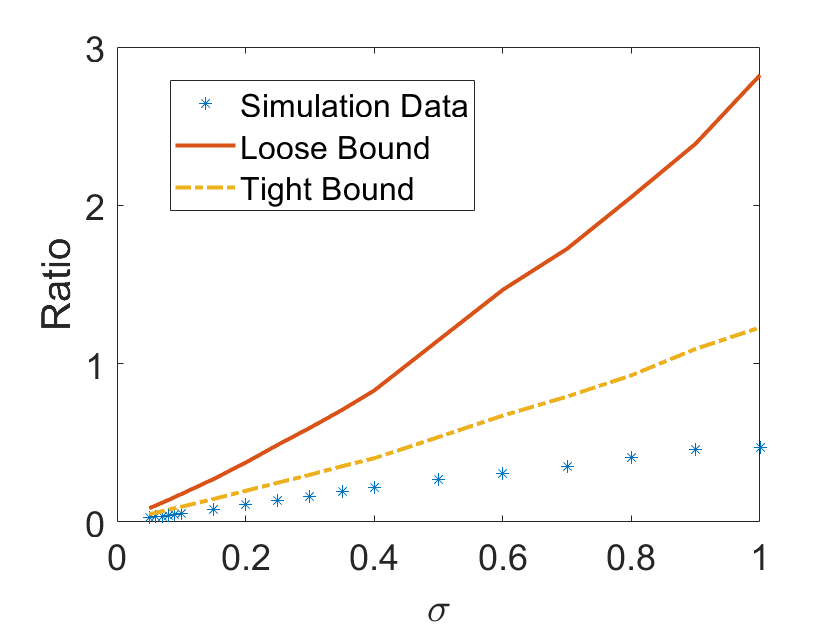}
\vspace{-6pt}
\caption{Estimation sensitivity to perturbation $\frac{\|\delta A_P\|}{\|A_P\|}$ and predicted bounds as noise magnitude increases.}
\label{error_bound}
\vspace{-6pt}
\end{figure}

\textbf{Toy Example:} 
Consider the simple linear system 
\begin{equation}
    \left[
          \begin{matrix}
            \xi_1\\
            \xi_2\\
          \end{matrix} 
          \right]_{k+1} = \left[
          \begin{matrix}
            1 & 2 \\
            3 & 4 \\
          \end{matrix} 
          \right]\left[
          \begin{matrix}
            \xi_1\\
            \xi_2\\
          \end{matrix} 
          \right]_{k}+\left[
          \begin{matrix}
            1 \\
            0.7 \\
          \end{matrix} 
          \right]u_k\enspace.
\end{equation}
We add zero-mean white noise to corrupt state observations, i.e. $N(0, \sigma^{2})$, with $\sigma\in [0.05,1]$ in $0.05$ increments. The initial condition is set at $[0.01,0.01]$. We perform a Monte Carlo simulation for $10,000$ trials to produce the average error. With reference to Fig.~\ref{error_bound}, the effect of the disturbance increases as variance increases (as expected). The general bound given in~\eqref{eq:loose} expands at higher rates than the tighter one we obtain via~\eqref{eq:tight} as the noise variance increases. 

\section{Control Structure Experiments}
We demonstrate the utility and evaluate the performance of DHC in simulating a planar quadrotor with different noise intensity, and then test experimentally with a quadrotor flying under the influence of ground effect.  

%

\subsection{Simulation in a Planar Quadrotor}
While quadrotor dynamics is well understood,
there is a certain limit on the degree of variations in parameters like mass and moment of inertia that model-based controllers can handle. Such variations make the utilized model inaccurate, which in turn may render the controller unstable. Online adjustment can help quantify model errors and inform the controller so as to maintain stable operation. 
\begin{figure}[!htbp]
\vspace{0pt}
\centering
\includegraphics[width = 0.21\textwidth]{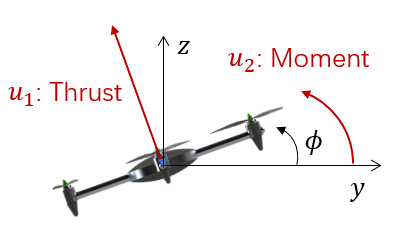}
\vspace{-9pt}
\caption{Planar quadrotor.}
\label{planar}
\vspace{-6pt}
\end{figure}

\subsubsection{System Description}
The planar quadrotor depicted in Fig.~\ref{planar} can be modeled as~\cite{sabatino2015model}
$$
  \begin{cases}
    m\ddot{y} = -u_1 \sin \phi\\
    m\ddot{z} = u_1 \cos\phi-mg\\
    I_{xx}\ddot{\phi} = u_2
\end{cases}.  
$$
Linearization around hover configurations yields  
\begin{equation}\label{eq:quad}
\begin{medsize}
    \dot{\xi} = \left[ \begin{matrix} & \textbf{0}_{3\times 3} & & \textbf{I}_{3\times 3} \\ {0} & {0} & {-g} & \\{0} & {0} & {0} & \textbf{0}_{3\times 3}\\{0} & {0} & {0} & \end{matrix}\right]\xi+\left[ \begin{matrix} &\textbf{0}_{4\times 2} \\ {1/m} &{0}\\{0} &{1/I_{xx}} \end{matrix}\right]u \enspace,
    \end{medsize}
\end{equation}
where $\xi = [y,z,\phi,\dot{y},\dot{z},\dot{\phi}]$ and $u=[u_1-mg,u_2]$. In all simulations, nominal parameter values match those of the physical Crazyflie robot used later for experiments, that is, $m = 0.03kg$, $I_{xx} = 1.43\times 10^{-5} kgm^2$, and $g = 9.8m/s^2$.

\subsubsection{Controller Design}
Besides our proposed DHC structure, we design and compare against a State Feedback Controller (SFC), a linear Model Predictive Controller (MPC), and a Model Reference Adaptive Controller (MRAC).
\begin{itemize}
    \item State Feedback Controller (SFC): With the closed-loop system poles set at [0.9;0.8;-0.9;-0.8;0.95;-0.95], we get 
\begin{equation*}
\begin{medsize}
    \hspace{-1pt}K_p\hspace{-2pt}=\hspace{-2pt}\left[\begin{matrix}   -0.0001 &  -0.0243 &  -0.0008 &   0.0012  &  0  &  0.0149\\
   -0.0000 & -0.0000 &  -0.0000 &  -0.0000 &   0  & -0.0000\end{matrix}\right]\hspace{-1pt}.
\end{medsize}
\end{equation*}

\item Model Predictive Control (MPC): A linear MPC is designed directly with model~\eqref{eq:quad} at hand. The prediction and control horizon is set to be $10$ and $2$, and the pitch angle $\phi$ is constrained in the interval $[-\pi/2,\pi/2]$.

\item Model Reference Adaptive Control (MRAC):
A parameter adaptation rule is used together with the previously described MPC to adjust the controller parameter so that the output of the plant tracks the output of the reference model having the same reference input.
The adaptation rule is designed based on \cite{mrac} as 
$$\frac{dm}{dt} = a_1(y_d-y)+b_1(z_d-z)$$
$$\frac{dI_{xx}}{dt} = a_2(y_d-y)+b_2(z_d-z)\enspace$$
where the weights were selected empirically to $a_1 = b_1 = 0.1$ and $a_2 = b_2 = 10^{-4}$. To tune the weights, 
we first chose their order to be consistent with the physical quantity they relate to. Then, we solved a sequential optimization problem to identify those parameter values that maximize MRAC's performance.
\item DHC: Our method also uses the linear MPC as the underlying controller. Then, we follow the steps in Algorithm~\ref{alg} to simulate the system. 
\end{itemize}

\subsubsection{Simulation}
The quadrotor's task is to reach position $[3;5]\;m$ from initial condition $[0;0]\;m$. We analyze the performance of each controller in two cases: when parameters are uncertain with low noise intensity (c1) where $\sigma_m = 0.2m$ and $\sigma_{I_{xx}} = 0.2 I_{xx}$, and with high noise intensity (c2) where $\sigma_m = 0.6m$ and $\sigma_{I_{xx}} = 0.6 I_{xx}$.
Perturbations in model parameters are generated at initialization of each simulated trial by sampling from normal distributions. Distribution means are the nominal values for $m$ and $I_{xx}$. Distribution variances are selected as above. Once the perturbation is sampled, it remains constant for the duration of the trial.\footnote{The specific selection of constant perturbations in model parameters emulates the case of inaccurate measurements of mass and inertia of the vehicle; these values may be approximately known (or estimated), but do not change during operation.  
The robustness properties of the proposed control structure to general perturbations is analyzed in Section~\ref{sec:sensitivity}.}
For every different condition we perform $10$ trials.

\begin{table*}[ht!]
\vspace{6pt}
\caption{Performance of different controllers in linear planar quadrotor model}
\vspace{-12pt}
\label{tab4}
\begin{center}
\begin{tabular}{|c|c|c|c|c|c|}
\hline
\multirow{2}{*}{Case} & \multirow{2}{*}{Controller} & Settling time & Total control effort to stabilize &  Steady state error & Overshoot\\&  &$T_s $ (s)  & $\Sigma|u|$& $|x_{desired}-x_{steady}|$  &$|x_{max}-x_{steady}|/|x_{steady}|$\\
\hline
\multirow{5}{*}{c1} & SFC & 1.95  & 1.2453&[0;0] & 2.5828\\
\cline{2-6}
 & MPC &  0.04  & 0.6020& [0.6846;0.7740]&0\\
\cline{2-6}
 & MRAC& 0.72  & 11.1852 &[ 0.2986;0.7038] & 1.4575\\
\cline{2-6}
 & DHC & 0.38  & 4.0847  & [0;0.0011]&0\\
\hline
\hline
\multirow{5}{*}{c2} & SFC &  Unstable&$\backslash$&$\backslash$ &$\backslash$\\
\cline{2-6}
 & MPC & 0.04   & 0.6032 & [1.6951;1.1906]&0\\
\cline{2-6}
 & MRAC & 0.92  & 11.0816 & [ 1.0351;1.0357]& 0.9718\\
\cline{2-6}
 & DHC & 0.38  & 3.6052& [0;0.0085]&0\\
\hline
\end{tabular}
\end{center}
\vspace{-18pt}
\end{table*}

\subsubsection{Results and Discussion}
Simulation results suggest that all controllers can stabilize the system in low noise intensity (see Table~\ref{tab4}). However, as noise intensity increases SFC fails. Linear MPC converges faster than all other methods, but produces increasingly large steady-state error. 
MRAC can reduce such steady-state errors, at the expense of longer settling time (within 2\%) and larger control effort, and could possibly cause a large overshoot as in SFC. 
Despite its efficiency shown here, tuning MRAC can be tedious, and its performance relies a lot on the tuned weights. 
On the other hand, our proposed DHC structure keeps updating the model that implicitly includes noise. As such, it can quickly extract refined dynamics from uncertain data to maintain system stability faster, with less control energy and smaller steady state error. Further, as the disturbance gets embedded in the learned model, the performance of controller does not deteriorate significantly as noise intensity increases.

\subsection{Controlling a Crazyflie Quadrotor Under Ground Effect}
We evaluate experimentally our DHC structure on a Crazyflie quadrotor operating under the influence of ground effect over sustained periods of time. 
Ground effect~\cite{c11} manifests itself as an increase in the generated rotor thrust given constant input rotor power~\cite{c35} when operating close to the ground or over other surfaces.  
When operating in ground effect, the dynamics of the robot-environment interaction change to a degree that an underlying model tuned for mid-air operation may no longer be valid~\cite{c37,kPOD}. In turn, model mismatches caused by varying robot-environment interaction (i.e. a form of uncertainty) degrade the robot's performance, both in terms of stability~\cite{c13} and control. 
As we show below, controllers unaware of ground effect will tend to raise the height of the robot when the latter flies sufficiently close over an obstacle. This behavior may lead to crashes when operating in confined and cluttered environments.  
Wrapping DHC around such a controller can help the robot adjust the impact of ground effect without any specific models for it. 

\subsubsection{Controller Design}
We follow the controller structure shown in Fig.~\ref{fig:MiC_cascade} (as in simulation). We test using two distinct lower-level controllers: a nonlinear geometric controller~\cite{c15}, and a PID controller.  
Controller gains are those pre-tuned by the manufacturer. Hence, note that the controllers have been tuned for operation in mid-air, not in near-ground operation. 
The controllers are different from those in simulation since 1) they are often used in practice, and 2) their underlying performance in simulation would depend on quality of tuning (especially for PID) and would thus not add significant value beyond the controllers already tested in simulation.

\begin{figure}[!htbp]
\vspace{-6pt}
\centering
\includegraphics[height=1.1cm]{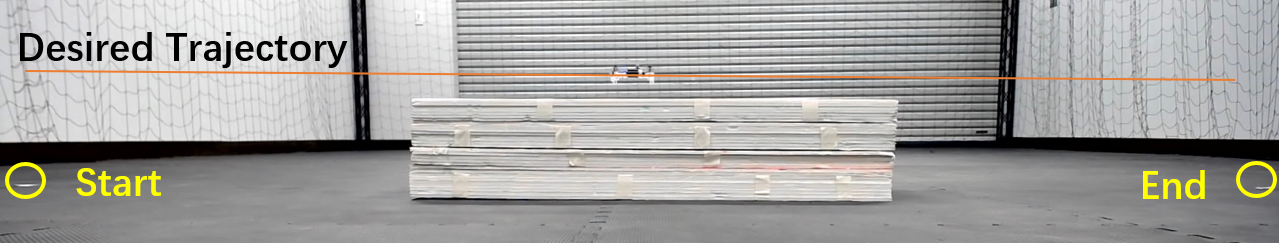}
\includegraphics[trim=5cm 5.6cm 5cm 5.6cm, clip, height=1.1cm]{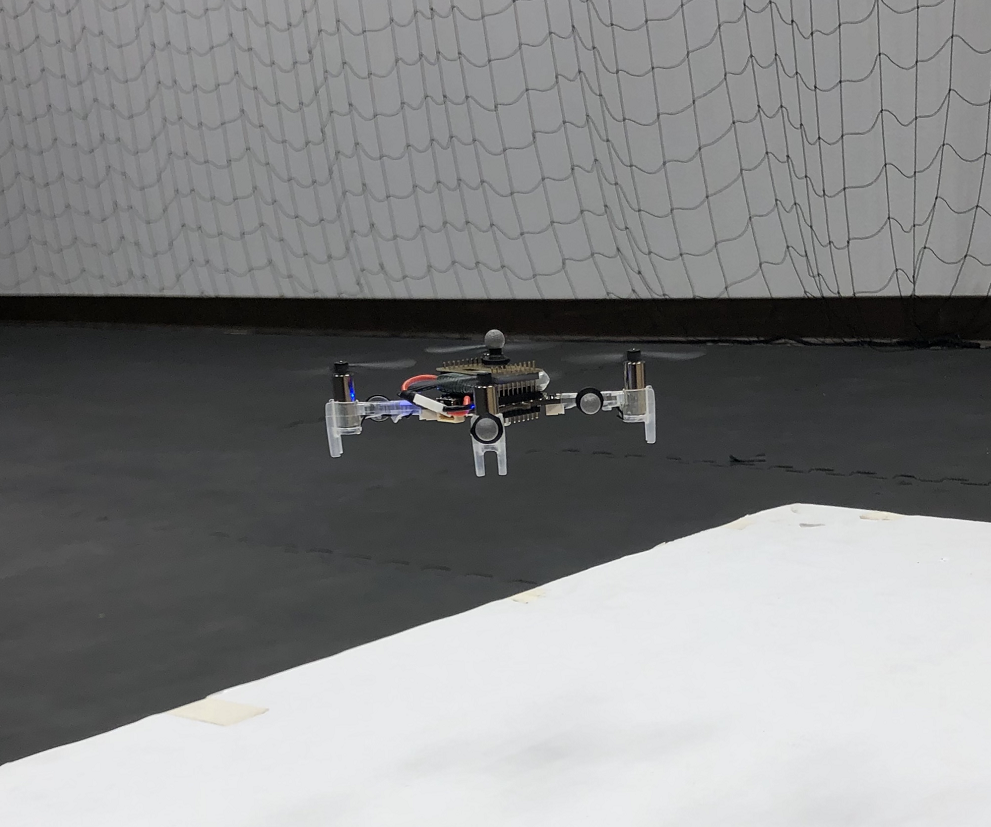}
\vspace{-6pt}
\caption{(Left) Experimental setup, and (Right) the Crazyflie. Sample experiments can be viewed at \url{https://youtu.be/OznDCskVnJU}.}
\label{quadrotor}
\vspace{-8pt}
\end{figure}

We evaluate the performance of both low-level controllers with and without DHC wrapped around them. 
The task is to fly at a constant height between an initial and a goal position (Fig.~\ref{quadrotor}). During part of the trajectory, the robot flies over an obstacle which creates sustained ground effect and influences robot behavior. 
Performance is quantified in terms of deviating from the desired height due to the ground effect.


\subsubsection{Experiment}
 
The robot is commanded to follow $2.4$\;m straight-line constant-height trajectories. The height of the obstacle is $h_o= 0.290$\;m. The robot propeller radius is $r= 0.023$\;m. We collect data at three distinct heights $h\in \lbrace h_o+r,h_o+2r,h_o+3r \rbrace$, and while flying at six distinct forward speeds $v\in \lbrace 0.4, 0.6, 0.8, 1.0, 1.2, 1.4 \rbrace$\;m/s. Experiments are run with the four aforementioned control structures. This leads to a total of $72$ distinct case studies. For each case study, we collect data from $10$ repeated trials. To minimize depleting battery effects, a fully charged battery is used at the beginning of each $10$-trial experimental session. Position data are collected via motion capture (VICON).

\subsubsection{Results and Discussion}
\begin{figure}[ht!]
\vspace{-9pt}
\centering
\includegraphics[trim= 2.2cm 2.1cm 3.2cm 1.5cm, clip,width =0.47\textwidth]{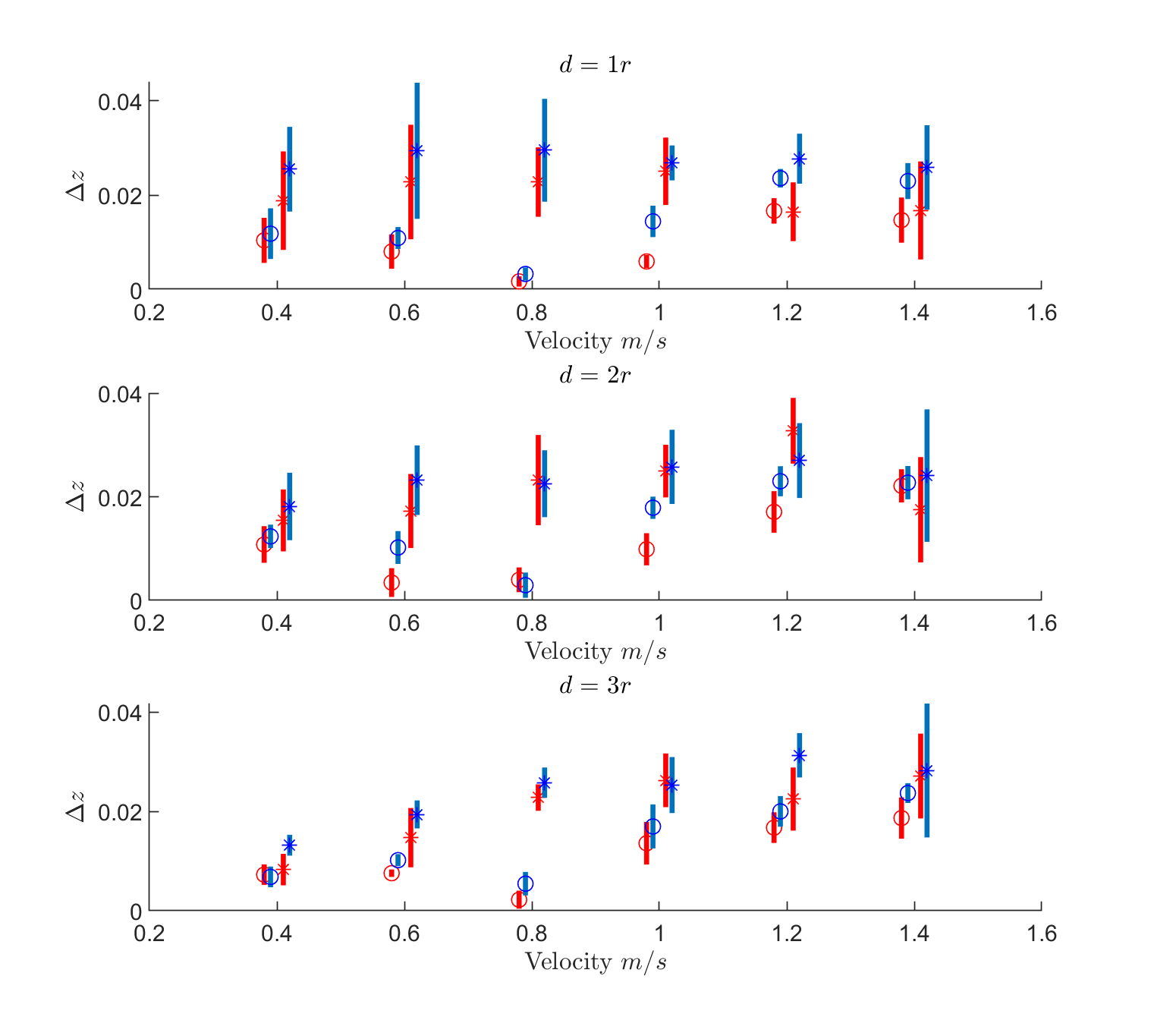}
\vspace{-6pt}
\caption{$1\;\sigma$ error plots for 
the standalone geometric controller (blue circle) and PID controller (blue star), and with wrapped DHC (red circle and star).}
\label{fig:analysis}
\vspace{-5pt}
\end{figure}

\begin{figure*}[!t]
\vspace{-5pt}
\centering
\includegraphics[trim= 3.5cm 2cm 2cm 2cm, clip,width = 0.98\textwidth]{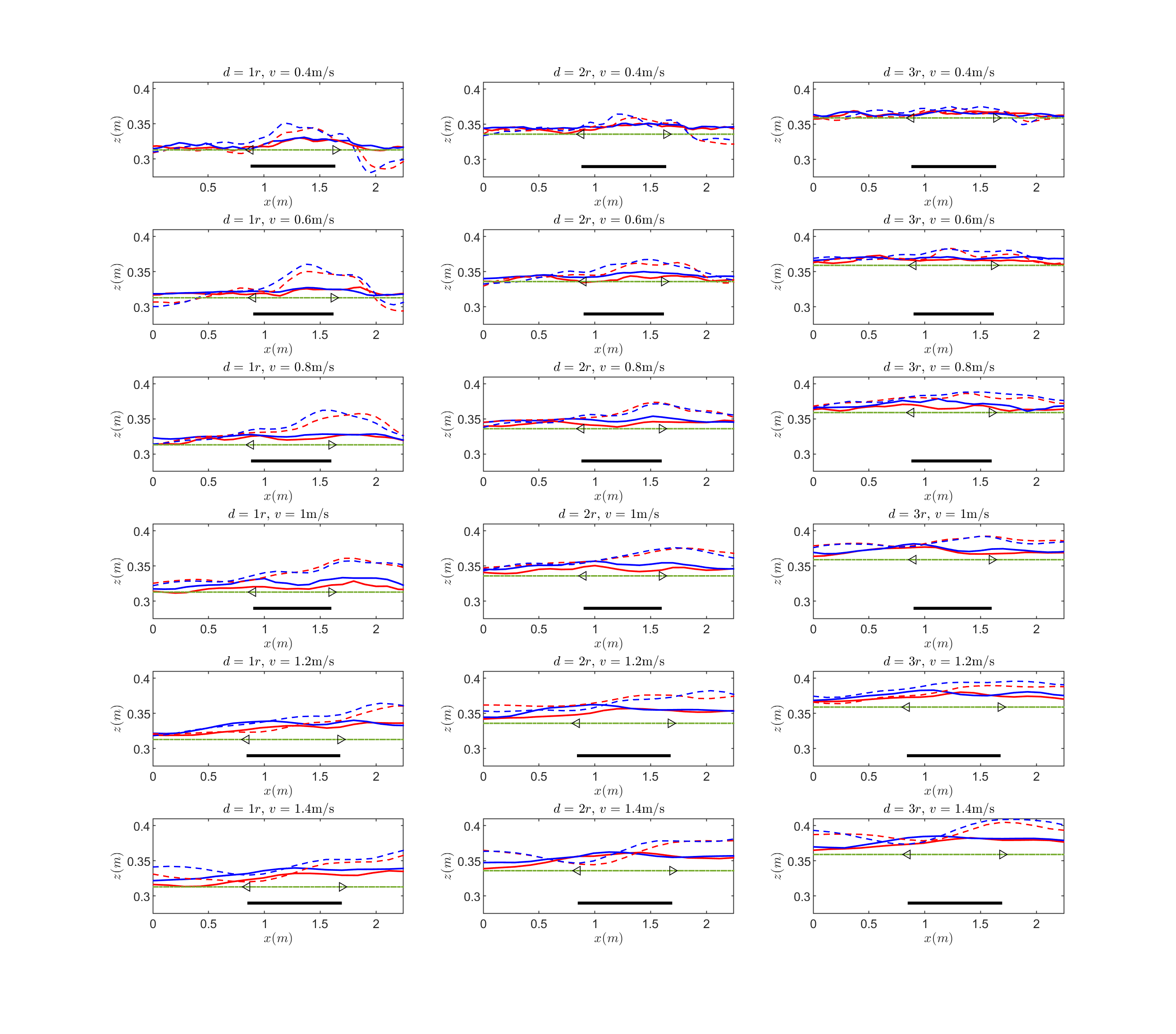}
\vspace{-30pt}
\caption{Trajectories of the Crazyflie tasked to fly at desired heights, $d$, over the obstacle (green dashed-dotted) and various forward speeds, $v$, under ground effect with standalone geometric controller (blue solid) and PID controller (blue dashed), and with wrapped DHC (red dashed and solid). Obstacle locations (black) are shown for clarity.}
\label{fig:ground_eff}
\vspace{-12pt}
\end{figure*}

Experimental results (Figs.~\ref{fig:analysis} and~\ref{fig:ground_eff}) reveal that wrapping the proposed hierarchical structure around a low-level controller (red curves in figures) can help keep the quadrotor closer to the desired height when compared to the standalone original controller (blue curves), on average. 
This indicates that incorporating model identification and controller adaptation modules through DHC helps the overall control structure adjust better and faster to uncertain perturbations as, in this case, ground effects. 

Furthermore, model/controller adaptation appears to perform better at low distances from the ground and as the forward velocity increases. This observation is in line with aerodynamics suggesting that 1) ground effect starts diminishing when the flying height above the obstacle increases to $d>3r$, and 2) as forward speed increases both ground effect and drag affect rotorcraft flight in still not well-understood ways~\cite{c35}. Hence, through this experiment we show how DHC can be used to render a controller adaptive to environment changes that excite unmodeled dynamics.

The hierarchical structure may still remain bound to the low-level controller's possible limitations, as shown in simulation. In the experiments, this is observed as forward speed increases. After clearing out the obstacle, the controller's apparent steady-state error seems to increase (c.f. bottom two rows of Fig.~\ref{fig:ground_eff}) at higher speeds.  While this may be an artefact of the limited experimental volume, the settling time of the controller with and without DHC still increases. Future work will investigate the interplay between the lower-level controller and DHC, and consider different hierarchical adaptive control structures to change an underlying controller's behavior more drastically when needed.



\section{Conclusions}
The paper introduced a Data-driven Hierarchical Control (DHC) structure with two key functions. 1) DHC utilizes an online data-driven approach to learn a model of input reference and true outputs. 2) DHC utilizes the derived model to keep refining a reference signal given to an underlying low-level controller to ensure system performance in the presence of system and/or system-environment uncertainty. We showed that DHC retains the stability properties of the underlying lower-level controller, and further determined lower bounds for our method's sensitivity to noisy data. 

The utility and performance of the proposed approach are tested in simulation and experimentally with a quadrotor. The test cases were designed to excite specific types of uncertainty that are common in practice: 1) constant model parameter errors in simulation; and 2) variable uncertain robot-environment interaction experimentally through operation under ground effect. Results suggest that DHC can successfully be wrapped around existing controllers in practice, to improve their performance by discovering and harnessing unmodeled dynamics during deployment. 


\bibliographystyle{IEEEtran}
\bibliography{IEEEabrv,IEEEexample}

\end{document}